\documentclass{vgtc}                          

\ifpdf
  \pdfoutput=1\relax                   
  \usepackage{graphicx}                
  \DeclareGraphicsExtensions{.pdf,.png,.jpg,.jpeg} 
\else
  \usepackage{graphicx}                
  \DeclareGraphicsExtensions{.eps}     
\fi%

\graphicspath{{figures/}{pictures/}{images/}{./}} 

\usepackage{microtype}                 
\PassOptionsToPackage{warn}{textcomp}  
\usepackage{textcomp}                  
\usepackage{mathptmx}                  
\usepackage{times}                     
\usepackage{cite}                      
\usepackage{tabu}                      
\usepackage{booktabs}    
\usepackage{colortbl}
\usepackage{algorithm}
\usepackage{algpseudocode}
\usepackage{csquotes}

\onlineid{1010}

\vgtccategory{Technique or Algorithm}

\newenvironment{tight_enumerate}{\begin{enumerate} \itemsep
-1.5pt}{\end{enumerate}}

\definecolor{teal}{RGB}{0, 125, 125}

\newcommand{\pheading}[1]{\vspace{3px}\noindent\textbf{#1}}
\begin{document}

\title{What \textit{Is} the Difference Between a Mountain and a Molehill?  Quantifying Semantic Labeling of Visual Features in Line Charts}

\author{Dennis Bromley\thanks{e-mail: dbromley@tableau.com}\\ %
        \scriptsize Tableau Research, Seattle, WA, USA %
\and Vidya Setlur\thanks{e-mail: vsetlur@tableau.com}\\ %
     \scriptsize Tableau Research, Palo Alto, CA, USA %
}

\keywords{Semantics, trends, annotation, text generation.}

\teaser{
  \centering
    \includegraphics[clip, trim=0cm 3.5cm 0cm 0cm, width=\textwidth]{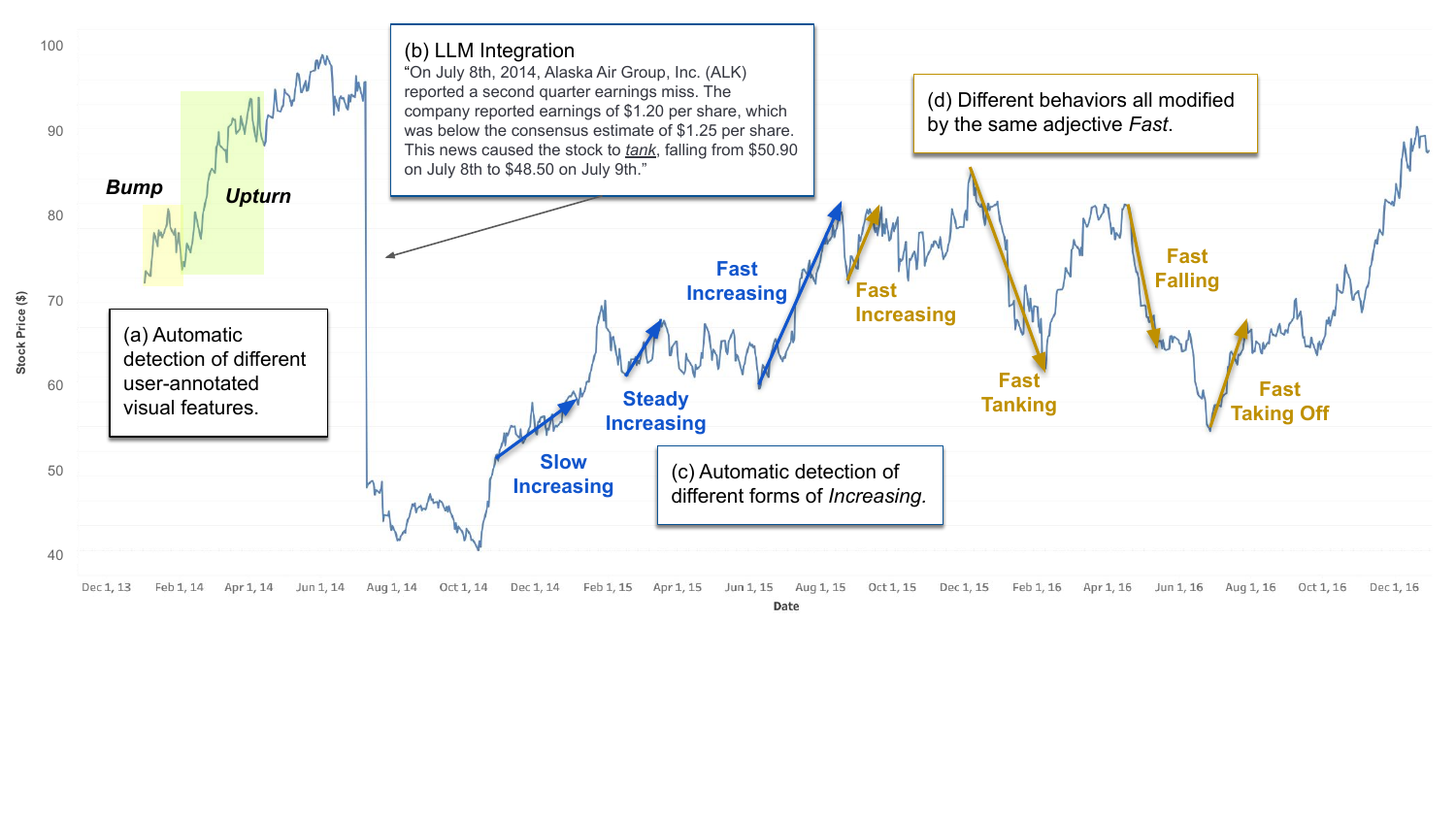}
    \caption{Automatic labeling of visual features in a line chart. (a) Automatic detection of user-annotated visual features. (b) Integration of discovered visual features with a large language model (LLM). (c, d) Automatic labeling of increases and decreases in trends using quantified semantics; the adjective/verb pairings encode specific empirically-derived line-slope information for generating corresponding annotations.}
  \label{fig:teaser}
}

\abstract{%
 Relevant language describing visual features in charts can be useful for authoring captions and summaries about the charts to help with readers' takeaways. To better understand the interplay between concepts that describe visual features and the semantic relationships among those concepts (e.g., \textit{`sharp increase'} vs. \textit{`gradual rise'}), we conducted a crowdsourced study to collect labels and visual feature pairs for univariate line charts. Using this crowdsourced dataset of labeled visual signatures, this paper proposes a novel method for labeling visual chart features based on combining feature-word distributions with the visual features and the data domain of the charts. These feature-word-topic models identify word associations with similar yet subtle differences in semantics, such as \textit{`flat,'} \textit{`plateau,'} and \textit{`stagnant,'} and descriptors of the visual features, such as \textit{`sharp increase,'} \textit{`slow climb,'} and \textit{`peak.'} Our feature-word-topic model is computed using both a quantified semantics approach and a signal processing-inspired least-errors shape-similarity approach.  We finally demonstrate the application of this dataset for annotating charts and generating textual data summaries.

}


\CCScatlist{
  \CCScatTwelve{Human-centered computing}{Visu\-al\-iza\-tion}{Visu\-al\-iza\-tion systems and tools}{};
}





\maketitle

\section{Introduction}

Data visualizations are often accompanied by captions and summaries describing their key takeaways to the reader~\cite{echeverria:2018,whitaker:2017}. Studies have also found that
captions help readers with takeaways by emphasizing visually prominent features in charts~\cite{kim2021towards}. There are well-established visualization techniques to draw a reader's attention to the visually prominent features of a chart using techniques such as highlighting and annotating a portion of the chart or changing the scale or data extents to show prominent visual patterns~\cite{cardbook}. 

However, the language for describing and emphasizing visual features in the captions is a less-studied topic. Automatic chart captioning and summarization tools~\cite{tableau,chen2019neural,cui2019datasite,hu2018dive} support authors with caption generation that describes visual features in charts; the language employed in the recommended captions and annotations tends to be simple, ranging from describing the domain, axes, and encodings to specific statistical information (e.g., extrema) describing specific marks in the chart~\cite{kim2021towards}. Little research has explored the nuances in language to add emphasis to the characteristics of the data encoded in a chart. \emph{Hedging} is a communicative strategy used in language for increasing or reducing the force of statements to emphasize and bring a reader's attention to specific portions of the text that is important to the overall takeaway of the intended message~\cite{hyland1998hedging}. Prior work in linguistics research indicates the benefits of employing hedge words to provide additional texture and emphasis to textual discourse for the reader~\cite{lakoff1973hedges}.

The goal of this work is to explore language that emphasizes visual features in charts and the semantic relationships that expose the nuances among those language concepts. We specifically focus on how people label and associate hedge words such as \textit{`sharp'} or \textit{`gradual'} when describing visual features in univariate line charts. To better understand the correspondence between language and visual features, we crowdsource a dataset of labeled visual chart features based on combining feature-word distributions with the visual features
and the data domain of the charts. These feature-word-topic models identify word associations with similar yet subtle differences in semantics, such as \textit{`flat,'} \textit{`plateau,'} and \textit{`stagnant,'} and descriptors of the visual features, such as \textit{`sharp increase,'} \textit{`slow climb,'} and \textit{`peak,'} for example. Our feature-word-topic model is computed using both a quantified semantics approach and a signal processing-inspired multi-resolution approach wherein windowed versions of crowdsource-labeled chart segments are applied to unlabeled charts to find regions of low-mean absolute error (MAE) shape similarity. We demonstrate the utility of this dataset for automatically annotating line charts and for generating data summaries using large language models (LLMs), as shown in Figure~\ref{fig:teaser}. We also describe future research directions for this work, such as incorporating domain-specific descriptions, leveraging LLMs for semantic enrichment, and supporting search and natural language (NL) interaction.




\section{Related Work}
This work builds on prior work that explores chart annotation techniques and linguistic approaches for generating data narratives.

\subsection{Text Annotation for Charts}
There is a growing body of research that focuses on the role and importance of text in visual analysis~\cite{stokesgive,ottley2019curious}. Kong et al.~\cite{kong2018frames,kong2019trust} evaluated how titles can impact the perceived message of a chart and found that people were more likely to recall information conveyed by slanted framings (e.g., emphasizing only part of a chart's message) than the actual chart's visuals. Kim et al. found that when both the chart and text described a high-prominence feature, readers treated the doubly emphasized high-prominence feature as the takeaway~\cite{kim2021towards}. When the text described a low-prominence chart feature, readers relied mostly on the chart alone and usually reported a higher-prominence feature as the takeaway. Hearst \& Tory examined participant preferences for text with visualizations in the context of chatbot interaction~\cite{hearst2019would}. Their study found that when participants preferred to see charts, they also preferred to view additional contextual data to be provided in the chart. 

Prior research has also explored how authors add annotations and descriptions to charts guiding a reader's attention to visual features in the chart, explaining what the underlying data means, and providing additional context~\cite{segel:2010,hullman:2011,Autotator:2019}. Kong and Agrawala developed techniques for analyzing charts to recover visually salient features of the data-encoding marks (e.g., min, max, mean values). Users can interactively add graphical and text annotations to facilitate chart reading~\cite{kong2012graphical,kong09}. Kandogan~\cite{kandogan:2012} introduced just-in-time descriptive analytics by employing statistics to automatically generate annotations for clusters and outliers. Contextifier~\cite{Hullman2013} uses news headlines to provide external contextual annotations for line charts. They consider linguistic relevance, the number of article views, and the visual saliency of chart peaks to identify the headlines and chart features to annotate. Henkin \& Turkay~\cite{henkin2020words} have done extensive work quantifying crowdsourced semantics for scatter plots.

Our work contributes to this body of work by exploring how hedge words can further express and describe visual features in line charts. We also quantify the semantics to identify language subtleties to automatically label and generate text summaries for describing the magnitude of the slopes and characteristics of the features.

\subsection{Linguistic Approaches for Generating Narratives }
The computational linguistics community has implemented techniques for identifying hedging patterns in text and conversational transcripts to determine their effectiveness in debating or communicating a point of view to the reader~\cite{islam-etal-2020-lexicon,goodluck:2021}. Other work has focused on creating datasets containing hedge cues, curated from open-access text, that are fed into a multitask learning model for text classification and generation~\cite{hedgepeer}. However, none of these linguistic approaches have explored hedging and its associated semantics for specifically describing visual features in charts.

Visual analytics systems incorporate generated text with visualization responses to help communicate key insights to the user~\cite{mittal1998describing, srinivasan2018augmenting, kanthara2022chart, chen2020figure, qian2021generating}. Other tools produce text summaries with statistical descriptions shown in the visualizations~\cite{tableau,mspbi2022}. Data storytelling incorporates textual narratives with visuals, communicating insights that are more memorable, persuasive, and engaging than statistics alone~\cite{segel:2010,kosara2013storytelling, lee2015more}. Systems like Kori~\cite{latif2019authoring, latif2021kori} and VizFlow \cite{sultanum2021leveraging} provide explicit linking strategies between text and charts to support design patterns for data storytelling, narrative sequencing, and rhetoric \cite{hullman2013deeper,hullman:2011}. In this paper, we further explore the interplay between text and charts by the automatic labeling of visual features in charts and text generation containing hedge word descriptors using a crowdsourced dataset of labeled visual signatures.

\section{Crowdsourcing Labeled Visual Features}
The motivation for crowdsourcing a labeled dataset of terms and visual features is two-fold: 1) capture semantic descriptions of different visual features in univariate line charts and 2) elucidate and quantify the relationships among those semantic descriptions.

    \begin{figure}[ht]
        \centering
        \includegraphics[width=\columnwidth, trim={0 0 8.5cm 0},clip]{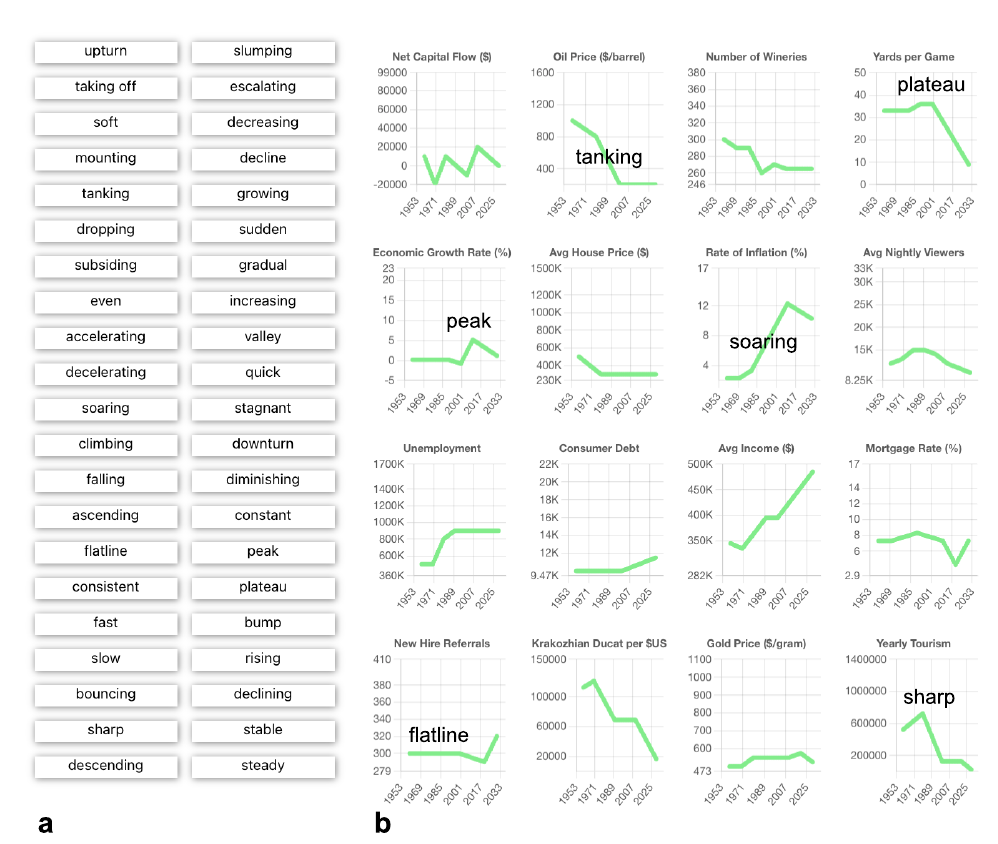}
        \caption{The annotation-collection tool. Participants drag words from the left (a) over to visual features of the charts on the right (b). The words are snapped to the nearest chart position. Words may be moved or deleted once they are attached to a chart. Individual words may be used on multiple charts and multiple times on a single chart.}
        \label{fig:datacollectiontool}
    \end{figure}

We collected labeled annotations for visual features in univariate line charts by implementing a data collection tool, shown in Figure~\ref{fig:datacollectiontool}. The tool was implemented as a Typescript frontend and a Django backend attached to a PostgreSQL database.

The interface has two parts: the left side (Figure \ref{fig:datacollectiontool}a) comprises $42$ word labels consisting of: 1) words related to the basic shape descriptors, \textit{`up,'} \textit{`down,'} and \textit{`flat,'} 2) adjectives that describe such shapes (e.g., \textit{`slow,'} \textit{`sudden,'}) and 3) words that describe the emergent shapes created by such regions (e.g., \textit{`plateau'} or \textit{`valley'}). To find these word descriptors, we leveraged the hierarchy of hypernyms and hyponyms from Wordnet~\cite{wordnet}, whose depth typically ranges up or down to two hierarchical levels (e.g., $`up' \rightarrow [`increasing,' `ascending']$), as well as word2vec~\cite{word2vec} to identify related concepts, such as \textit{`sharp'} and \textit{`increasing.'} In total, the list contained 8 nouns, 13 adjectives, and 21 verbs. Note that while this list is not exhaustive, we considered the set of words as a starting point for collecting nuanced language that describes common features found in line charts. 
The words were displayed in a randomized order in the interface to avoid positional bias.

The right side of the tool interface (Figure \ref{fig:datacollectiontool}b) displayed $16$ line charts shown in random order to each participant to mitigate any positional bias. The same charts were shown to all participants. The charts were generated in Chart.JS~\cite{chartjs}, showing years on the x-axis, ranging from 1960 to 2030. The title and its corresponding y-axis range were randomly assigned from one of the following topics: Average Income (\$), Unemployment, Yards per Game, New Hire Referrals, Yearly Tourism, Rate of Inflation (\%), Average House Price (\$), Krakozhian Ducats per \$US, Average Nightly Viewers, Economic Growth Rate (\%), Gold Price (\$/gram), Oil Price (\$/barrel), Consumer Debt, Number of Wineries, Mortgage Rate (\%), and Net Capital Flow (\$). Each chart is a line graph constructed by connecting seven sequential line segments end to end.  Similar to the chart stimuli generated in~\cite{kim2021towards}, each segment is randomly assigned one of nine different slopes: \textit{Up}, \textit{Down}, \textit{Flat} with slopes [1, -1, 0],
\textit{SteepUp}, \textit{SteepDown}, \textit{SteepFlat} with slopes [3, -3, 0]
 \textit{GentleUp}, \textit{GentleDown}, \textit{GentleFlat} with slopes [0.5, -0.5, 0].
 
We recruited $67$ participants through a mailing list at a data analytics software company. Participants were required to pass a chart literacy test before proceeding to the annotation labeling exercise. Participants annotated the charts by dragging words from the left (Figure~\ref{fig:datacollectiontool}a) onto the charts on the right (Figure~\ref{fig:datacollectiontool}b) in the interface. Multiple words could be dragged to the same feature in a chart. We recorded the chart identifier, the annotation, the position along the line graph where the annotation occurred, the date the annotation occurred, and a unique anonymous participant identifier. The study details and instructions are found in the supplementary material.

\section{Analysis of the Dataset}
\subsection{Analysis Technique}
We calculate term co-occurrence and perform annotation clustering to identify quantifiable relationships among the different annotation terms. Annotation co-occurrence helps us understand how often different annotation terms are used to label the same visual feature; for each annotation, the co-occurrence of every other word is calculated as the average of per-segment \% representation. For example, consider two segments that contain the annotation \textit{`quick.'} If the term \textit{`fast'} represents 50\% of the annotations on the first segment and 30\% of the annotations on the second segment, then the overall co-occurrence of \textit{`fast'} with respect to \textit{`quick'} is $\frac{50\% + 30\%}{2} = 40\%$. Note that co-occurrence is not symmetric as \textit{`quick'} may co-occur with different annotations than \textit{`fast.'} 

Annotations are clustered using hierarchical clustering and Ward's linkage~\cite{ward1963hierarchical} calculated with Euclidean distance; these approaches tend to identify dense clusters while making a minimum number of assumptions about cluster size, shape, and count. Position matrix entries are assigned by segment co-occurrence. For example, if \textit{`quick'} and \textit{`fast'} co-occurred 10 times, then each would have the position $10$ on the other's axis. The matrix is then scaled so all values are in $[0, 1]$, and values of $1.0$ are placed along the diagonal.

\subsection{Findings}
\begin{figure}
    \centering
    \includegraphics[width=0.5\textwidth]{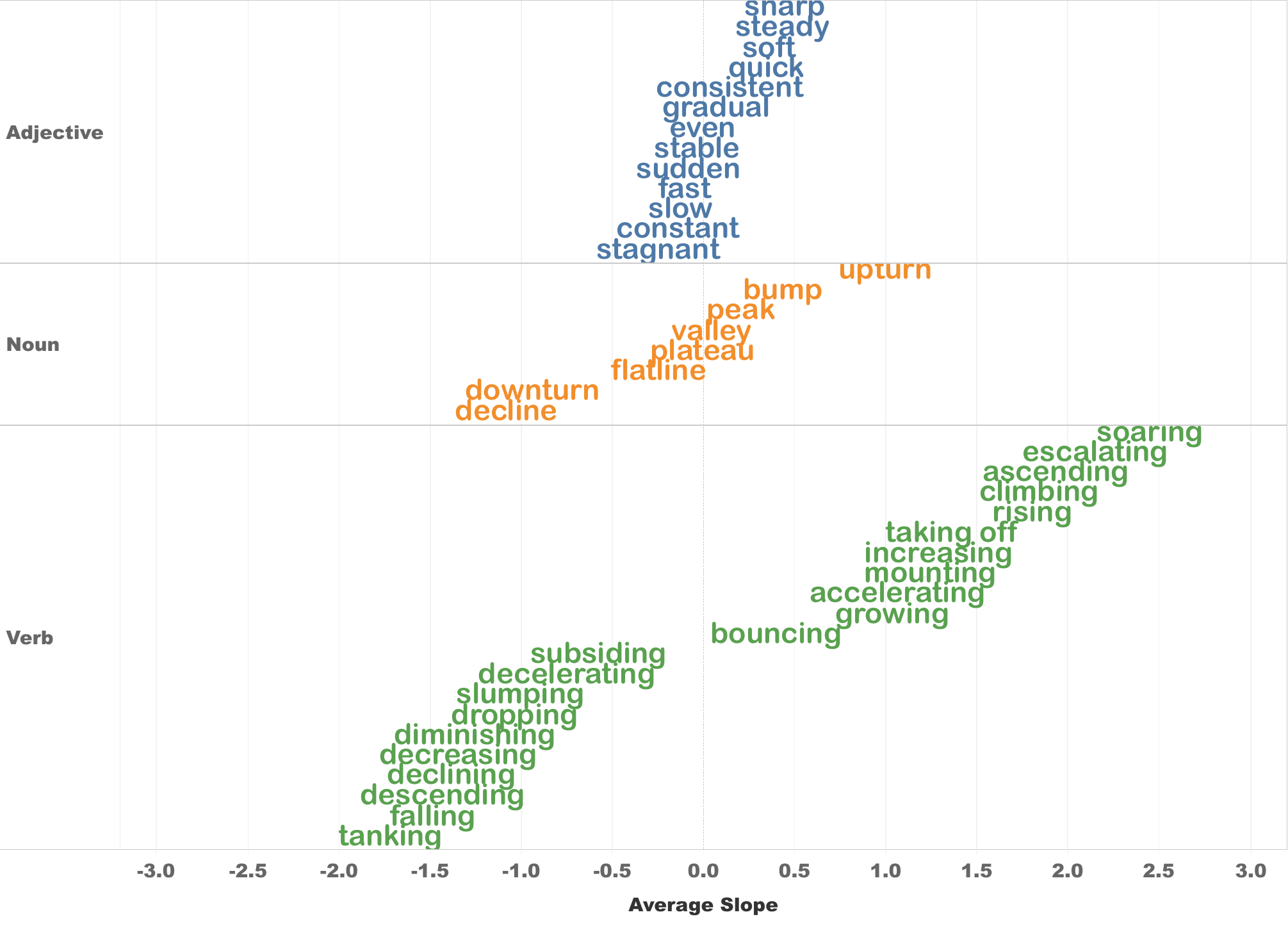}
    \caption{Average segment slope of the annotations.  The maximum possible slope range available for the charts is -3 to +3.}
    \label{fig:averageslope}
\end{figure}

A total of $67$ participants generated $1,892$ annotations, with an average of $28.2$ annotations per user, and $118.3$ annotations per chart.  $7$ segments and $16$ charts provided a total of $112$ different segments. On average, there were $17$ annotations per segment, allowing us to empirically derive various inter-word relationships (Figure \ref{fig:averageslope}). Term co-occurrence analysis quantifies which words are typically present together. Agglomerative hierarchical clustering of term co-occurrence results in distinct groups, suggesting a high degree of semantic agreement among participants. The crowdsourced dataset and analysis are provided in the supplementary material.

One of the goals of this work is to understand the hierarchical semantics of the visual feature/annotation pairs. Using line slope as a fundamental component of signal shape, we analyzed the average slope associated with each annotation. As shown in Figure \ref{fig:averageslope}, slope analysis distributed the various annotation words across a broad continuum from steeply descending to steeply ascending. Not only does this suggest an empirically derived semantic hierarchy e.g., \textit{`soaring'} is a steeper increase than \textit{`taking off'} while \textit{`tanking'} is a steeper dropoff than \textit{`slumping,'} the quantification of that hierarchy allows us to make concrete NL recommendations when generating labels for previously unlabeled signals (Figures \ref{fig:teaser}c and \ref{fig:teaser}d).

\section{Applications}
We demonstrate two applications for this crowdsourced dataset of labeled visual signatures. 

\subsection{Automatic Labeling of Visual Features in Line Charts}
Using the dataset, we develop two techniques for automatic labeling of visual features: \textit{shape identification}, which is useful for discovering concrete shapes such as \textit{`peak'} or \textit{`valley'}, and \textit{slope identification}, which is useful for describing how univariate data changes along the y-axis. In this discussion, we borrow from the signal processing lexicon and refer to a univariate data set as a \textit{signal} and the small annotated source signal whose shape were are looking for in a larger unlabeled signal as the \textit{kernel}.

\textit{Shape identification} tries to find an annotated visual feature in a larger unlabeled signal. Figure \ref{fig:teaser}a shows the detection of a \textit{`bump'} and an \textit{`upturn.'} This shape identification approach is particularly applicable to finding visual features that are constructed from multiple segments, e.g., a \textit{`peak'} consists of a rising segment followed by a falling segment. The following algorithm describes the process of identifying a kernel signal’s shape within a larger unlabeled signal.

\subsubsection{Shape Discovery Algorithm}
\label{supp:algorithm_annotations}
\begin{tight_enumerate}
    \item Begin with an unlabeled signal in which we would like to find a visual feature.   
    \item Collect all 112 (16 line plots * 7 segments/plot) annotatable segments and the annotations associated with them.  
    \item For each segment, build a five-segment \textit{kernel signal} consisting of the annotated segment and two segments on each side of that annotated segment. Note that kernels near the edge may consist of fewer than five segments. For each such kernel signal, create shallow and deep variants of it where the normalized variant heights range from $[0.1, 1.0]$ in units of $0.1$. 
    \item For each variant, perform normalization, smoothing, and take the first derivative. We employ standard Savitzky-Golay smoothing~\cite{savitzky1964smoothing} with a smoothing factor proportional to the kernel size for efficient smoothing and derivation. 
    \item Similarly, normalize the unlabeled signal and apply Savitzky-Golay smoothing and derivation.
    \item Calculate a windowed-mean-absolute-error (MAE) by sliding the kernel past the unlabeled signal, much like convolution. The window size is parameterizable to allow the algorithm to search for visual features of different sizes. 
    \item Accrue these errors for every variant of every kernel.
    \item For every kernel, calculate MAE z-scores.
    \item Filter MAE scores using two criteria: max acceptable MAE score and z-score. Keep points below either threshold.
    \item Mark points that meet the criteria threshold. The presence of points indicates that a visual feature is found in the chart.
    \item Merge neighboring (<= 2 points) qualifying points into larger annotated regions.
\end{tight_enumerate}

In addition to the least-errors shape identification approach taken above, the quantified slope semantics shown in Figure \ref{fig:averageslope} provides us with an additional tool for visual feature identification. Specifically, the quantified slope semantics helps identify specific relationships among line slope, hedge words, and the hedge word's semantic modifiers.  For example, Figure \ref{fig:averageslope} shows us a rough hierarchy of single-word slope descriptions from which we might decide to label a line as \textit{`soaring'} rather than \textit{`growing.'} However, if we look at verb annotations \textit{and their adjective modifiers} as a single unit that encodes line-slope information, we become much more precise; for example, \textit{`taking off'} in the context of \textit{`gradual'} has an average slope of only 0.7, but \textit{`taking off'} in the context of \textit{`quick'} has a much steeper average slope of $2.7$. Using this information along with word co-occurrence data for the specific \textit{<adjective><verb>} pairs, we are able to annotate the different regions of the analyzed signals (Figures \ref{fig:teaser}c and \ref{fig:teaser}d) (refer to supplementary materials for expanded \textit{<adjective><verb>} data). Selecting an \textit{<adjective><verb>} annotation for a given chart region uses the following protocol:
\begin{tight_enumerate}
    \item Determine the slope of a given region using Ramer-Douglas-Peuker piecewise-linear decomposition~\cite{douglas1973algorithms}.
    \item Find all \textit{<adjective><verb>} pairs whose average slope falls within a window (default = 0.5) of the desired slope.
    \item From that set, select the \textit{<adjective><verb>} pair with the highest annotation co-occurrence. The window in step 2 allows us to use annotation co-occurrence to select more common expressions like \textit{`fast tanking'} instead of \textit{`stagnant accelerating.'}
\end{tight_enumerate}

One of the goals of this work is to quantify relationships between visual features and annotations. Figures \ref{fig:teaser}c and \ref{fig:teaser}d and Figure \ref{fig:averageslope} show that these terms do, in fact, work together to encode specific slope information that can be used to automatically annotate a univariate signal.  Among terms, annotation clustering (refer to supplementary data) shows that terms tend to cluster in semantically intuitive ways.  Collectively, these findings support the hypothesis that quantitative analysis of semantic labels may be capable of generating visual feature labels that are not only human-accessible but also quantitatively accurate. Providing language descriptors for accurately describing data insights can provide useful `guard rails' and guidance as data summary generation becomes prevalent with the use of LLMs.

\subsection{Visual Feature Integration with LLMs}
To leverage the generative language of LLMs, we combine the semantic labels with additional information from the data set to form input prompts; for example, we employ the stock symbol and the dates of the discovered visual feature to ask the GPT 3.5 LLM~\cite{brown2020language} the templated question, ``What happened between <\textit{July 8, 2014}> and <\textit{July 9, 2014}> that caused the stock symbol <\textit{ALK}> to <\textit{\textbf{tank}}>?". (Refer to supplementary materials for the prompt template.) 

The specific LLM response is shown in Figure \ref{fig:teaser}b.  Notice that the model's responses implicitly integrate additional data into the user's investigation. For example, no data involving share price, earnings reports, or even the company name is explicitly linked to our chart. While these results are preliminary, additional research needs to explore the effectiveness of LLMs as ad hoc data sources.

\section{Future Directions}
The crowdsourced data has several layers of contribution.  At a high level, simple shape identification enables the labeling of charts with appropriate language, whether for colloquial phrasing or domain-specific terminology.  At a deeper level, we have begun to quantify the relationships between visual features and various hedge words, suggesting the possibility of numerically-accurate NL data descriptions. Finally, the data shows promise for analyzing the relationships among the different hedge words. While we believe that our work is an initial step in exploring the interplay of language and visual features, we identify the following future research directions:


\pheading{Incorporate additional charts and domain-specific descriptions.} Our dataset currently applies to univariate line charts. Future work should investigate language descriptors for other chart types, as well as labels for describing concepts for specific domains, such as associating \textit{`flat'} with sales trends and \textit{`constant'} with temperatures~\cite{lakoff1973hedges}.

\pheading{Use of LLMs to further semantic enrichment.} We explored the use of GPT for summary generation using the labels describing visual features in a chart. The models do have limitations around higher-order numeracy reasoning and context~\cite{frieder2023mathematical}. For example, in Figure \ref{fig:teaser}b, while the LLM provided several reasons for the stock price decline, the model missed the fact that there was a stock split. Custom-trained GPT models could potentially bridge this gap in higher-order analytical reasoning by incorporating additional knowledge. Other utilities for these custom LLMs could explore the automatic enrichment of additional descriptors for the dataset. 

\pheading{Supporting the search of shape descriptors.} Annotations and summaries describing visual features in charts could be used as metadata in search interfaces to find pre-authored charts based on search queries such as, ``find me the sales chart that has a spike in 2009, followed by a gradual decline,'' or in a voice assistant to ask for real-time notifications about data - ``Hey Siri, tell me if this stock \textit{tanks}.'' The work could also provide language prompts to LLMs to support sketching interfaces used for generating data stories~\cite{chung2022talebrush}. 

\section{Conclusion}
This work explores the interplay of language and hedge words that describe visual features and their semantic relationships in line charts. We conducted a crowdsourced study to collect a range of label and visual feature pairs for these charts. Using this dataset of labeled visual signatures, we demonstrated its application for labeling charts and generating text summaries. The quantitative semantics presented in this work suggest a path forward for converting the crowdsourced dataset of feature-word descriptions into a semantic library of concepts that can distinguish between a \textit{`rise'} and a \textit{`gradual increase,'} for example. By making this dataset available to the broader research community, we believe that the work has useful implications for labeling and summarizing concepts for other chart types and their features, as well as for specific data domains. Our work suggests that, for the most part, people have a shared sense of semantic meaning. While the common saying, ``Don't make a mountain out of a molehill,'' reprimands the exaggeration of a minor issue, perhaps exploring the \emph{actual} difference between a mountain and a molehill is an important step towards better language and data understanding.

\newpage
\bibliographystyle{abbrv-doi}
\bibliography{main}

\end{document}



\maketitle

 These are the supplementary materials for the paper "What Is the Difference Between a Mountain and a Molehill? Quantifying Semantic Labeling of Visual Features in Line Charts".

\section{Quantitative Validation of Generated Charts}
\label{supp:quant_chart_validation}
    To confirm that the generated charts were broadly representative of the larger combinatoric possibilities, we measured factors that contributed to line shape and validated that those factors were distributed across the possible domains (Figure \ref{fig:chartmetrics}).
    
    The charts showed a broad distribution of up-trending, flat, and down-trending univariate lines.  Individual line segments within each univariate line were also well distributed with charts containing from 0-70\% of positive, flat, or negative slopes.  Finally, we wanted to present the participants with a range of simple and complex curves.  We quantified complexity by the number of inflection points in the curve; the more inflection points, the more complex the curve.  The distribution of inflection points for our curve set was well-distributed between 1 and 5 (0 inflections being an unchanging line, 6 inflections being the most complex with a slope change every segment), indicating that we had a distribution of both simple and complex curves.

    \begin{figure}
        \centering
        \includegraphics[width=0.5\textwidth]{figs/net_slopes.pdf}
        \includegraphics[width=0.5\textwidth]{figs/up_down_flat_histograms.pdf}
        \includegraphics[width=0.5\textwidth]{figs/inflection_points.pdf}
        \caption{Distributions of various chart-generation metrics.  Top: Overall chart slope ranged between slopes of -2 to 2.  The steepest possible curve range was -3 to 3.  Middle: Charts contained from 0-70\% of positive, flat, and negative slopes.  Bottom: Curve complexity ranged from 1-5 directional changes per curve.  }
        \label{fig:chartmetrics}
    \end{figure}

\section{Annotation Analysis}
\subsection{Annotation Saturation}
    As shown in Figure \ref{fig:annotation_saturation}, our data collection seemed to achieve a degree of saturation.

    \begin{figure}
        \centering
        \includegraphics[width=0.5\textwidth]{figs/Saturation graph.pdf}
        \caption{Unique annotation:visual-feature annotation entries over time.  Over the roughly 24 hour span in which participants annotated charts, approximately 70\% of the unique entries were entered in the first two hours.  The decrease in unique entries over time suggests a degree of saturation in our data collection.}
        \label{fig:annotation_saturation}
    \end{figure}

\section{Quantified Semantics}
\subsection{Agreement and Purity Factors}
\label{supp:agreement_and_purity_factors}

 To quantify the degree to which participants agreed on the annotation of visual features, we established an \textit{agreement factor} for each word:
    
 \[Agreement Factor = 1.0 - \frac{\textnormal{Number Of Annotated Segments}}{\textnormal{Number Of Annotations}}\]
        
Agreement factors range between zero and one; if 100 annotations of \textit{Hill} were spread over 100 different segments, the agreement factor would be $1.0 - 100/100 = 1.0 - 1.0 = 0$.  If the same 100 annotations of \textit{Hill} were concentrated on only a single segment, the agreement factor would be $1.0 - 1/100 = 1.0 - 0.01 = 0.99$.

To quantify the degree to which a particular word \textit{uniquely} describes a visual feature, we established a \textit{purity factor} for each word:
    
\[Purity Factor = Average(\frac{\textnormal{COUNT(Word) on This Segment}}{\textnormal{COUNT(All Words) on This Segment}})\]
        
Purity factors range between zero and one, where high values indicate that an annotation is used to uniquely identify visual features, and low values indicate that an annotation is only one of many labels for a given visual feature.


The agreement factor calculations indicate that participants often agreed on which visual features would be appropriately described by which annotations.  For example, people tended to agree on what \textit{peak}, \textit{bouncing}, \textit{valley}, \textit{tanking}, and \textit{soaring} meant.  The purity factor calculations, on the other hand, indicate a collective degree of confidence that a visual feature was \textit{uniquely} described by a given annotation; for example, participants tended to agree that the visual features described by \textit{flatline} and \textit{peak} could be uniquely labeled.  Agreement and purity factors are shown in Figures \ref{fig:agreementfactor} and \ref{fig:purityfactor}.

\begin{figure}
    \begin{subfigure}[t]{0.5\textwidth}
        \includegraphics[width=\textwidth]{figs/Agreement Factor of Words.pdf}
        \caption{Agreement factor}
        \label{fig:agreementfactor}
    \end{subfigure}
    \hfill
    \begin{subfigure}[t]{0.5\textwidth}
        \includegraphics[width=\textwidth]{figs/Annotation Purity Factor.pdf}
        \caption{Purity factor}
        \label{fig:purityfactor}
    \end{subfigure}
    \label{fig:factors}
    \caption{Agreement and Purity Factors}
\end{figure}

\subsection{Annotation co-occurrence and term clustering}
Annotation term co-occurrence and clustering is shown in Figure \ref{fig:association_and_clustering}.

\begin{figure}
    \begin{subfigure}[t]{0.265\textwidth}
        \includegraphics[width=\textwidth]{figs/Annotation Co-presence.pdf}
        \caption{Annotation Co-occurrence. The color indicates what percentage of co-occurring annotations belonged to a particular word.  Only the top 10\% of co-occurring words are shown here.  Note the red color of \textit{peak} indicating that it primarily co-occurs with itself.  This is consistent with it's high purity score in Figure \ref{fig:purityfactor}.}
        \label{fig:annotationcooccurrence}
    \end{subfigure}
    \hfill
    \begin{subfigure}[t]{0.2\textwidth}
        \includegraphics[width=\textwidth]{figs/annotation denrogram.pdf}
        \caption{Term Clustering.  Note the clean semantic segmentation among the three lower clusters.}
        \label{fig:annotationdendrogram}
    \end{subfigure}
    \label{fig:wordassociation}
    \caption{Annotation Association and Clustering}
\label{fig:association_and_clustering}
\end{figure}

\section{LLM Interaction}
\subsection{LLM prompt for Visual Feature Description}
\label{supp:llm_prompt_description}
The following prompt was used to query the Chat GPT 3.5 LLM to obtain information about an identified visual feature.
    \begin{displayquote}
    What happened between <\textit{Beginning Date}> and <\textit{Ending Date}> that caused the stock symbol <\textit{Stock Symbol}> to <\textit{Semantic Label}>?
    \end{displayquote}

\subsection{LLM prompt for External Data Sets}
\label{supp:llm_prompt_explicit_data}
The following prompt was used to query the Chat GPT 3.5 LLM for links to external data sources related to the identified visual feature. 
    \begin{displayquote}
    Give me a json formatted list of hyperlinked internet resources that might be related to the stock <\textit{Stock Symbol}> price <\textit{Semantic Label}> between <\textit{Beginning Date}> and <\textit{Ending Date}>.  Include any public datasets, news sources, social media, and industry publications that are relevant to the industry.  The JSON should for formatted as a list where each item has a 'name' member and a 'url' member.
    \end{displayquote}

The LLM response for the prompt is shown in Listing \ref{gptjson}.  As discussed in the paper, many of these links are in some way erroneous, consistent with the current recognition that LLMs, while powerful, still have work to do to improve their correctness.
\label{supp:llm_explicit_data_response}
    \begin{listing}
            \begin{minted}
            [
            frame=lines,
            framesep=2mm,
            baselinestretch=1.2,
            bgcolor=White,
            fontsize=\tiny,
            ]
            {js}
    [
        {
            "name": "AMGN Stock Price History",
            "url": "https://finance.yahoo.com/quote/AMGN/history?p=AMGN"
        },
        {
            "name": "AMGN Twitter Feed",
            "url": "https://twitter.com/search?q=%40amgn&src=typd"
        },
        {
            "name": "AMGN Investor Relations",
            "url": "https://investor.amgen.com/"
        },
        {
            "name": "AMGN SEC Filings",
            "url": "https://www.sec.gov/cgi-bin/browse-edgar?CIK=AMGN&owner=exclude&action=getcompany"
        },
        {
            "name": "AMGN Industry Reports",
            "url": "https://www.marketwatch.com/investing/stock/amgn/industry"
        }
    ]    
            \end{minted}
        \caption{JSON-formatted external data resources recommended by the GPT 3.5 LLM.} 
        \label{gptjson}
    \end{listing}

\section{Word Co-Occurrence Among Different Parts of Speech}
\label{supp:annotation_coocurrence}
    The collected data in Figure \ref{fig:word_cooccurrence_extended} indicate that we can quantify the effect of a descriptive word on another word e.g. an adjective or adverb on a verb.  Collectively, the words encode information about the slope of the geometry, suggesting that semantic nuance and numerical accuracy may be able to go hand in hand.

    \begin{figure}
        \centering
        \includegraphics[width=1.2\textwidth, angle=90]{figs/Annotation Co-Presence Extended.pdf}
        \caption{Annotation co-occurrence.  This figure shows the average slope of the top 50\% of word co-occurrences.  Note that co-occurrence is not symmetric; \textit{tanking}'s average slope value in the presence of \textit{quick} (-2.2, upper right) is not the same as \textit{quick}'s average slope in the presence of \textit{tanking} (which does not even make the 50\% cutoff).  Color blocks delineate the intersections among the different parts of speech.}
        \label{fig:word_cooccurrence_extended}
    \end{figure}